\documentstyle[amsfonts,epsfig,12pt]{article}
\newcommand{\be}{\begin{equation}}
\newcommand{\ee}{\end{equation}}
\newcommand{\bea}{\begin{eqnarray}}
\newcommand{\eea}{\end{eqnarray}}
\newcommand{\beas}{\begin{eqnarray*}}
\newcommand{\eeas}{\end{eqnarray*}}
\newcommand{\bi}{\begin{itemize}}
\newcommand{\ei}{\end{itemize}}
\newcommand{\bc}{\begin{center}}
\newcommand{\ec}{\end{center}}
\newcommand{\bfl}{\begin{flushleft}}
\newcommand{\efl}{\end{flushleft}}
\newcommand{\bfr}{\begin{flushright}}
\newcommand{\efr}{\end{flushright}}
\newcommand{\f}{\frac}

\def\6{\partial} \def\a{\alpha} \def\b{\beta}
\def\g{\gamma}  
\def\e{\epsilon}
  
  \def\l{\lambda}
\def\m{\mu}   
\def\r{\rho} \def\s{\sigma} \def\t{\tau}
  
\def\o{\omega} \def\G{\Gamma} 
  
  \def\O{\Omega}

\newcommand{\HH}{{\cal H}}

\begin{document}
\title{Thermal String Vacuum in Black-Hole $AdS$ Spacetime}
\author{E. L. Gra\c{c}a$^{ab}$\footnote{egraca@ufrrj.br}
and
I. V. Vancea$^{c}$\footnote{ion@dfm.ffclrp.usp.br}}
\date{May 10, 2005}
\maketitle

\begin{center}
{\small 
${}^a${\em Departamento de F\'{\i}sica, Universidade Federal Rural do Rio de Janeiro},\\
{\em BR 465-07-Serop\'{e}dica, RJ, Brasil}\\
${}^b${\em Centro Brasileiro de Pesquisas F\'{\i}sicas},\\
{\em Rua Dr. Xavier Sigaud 150, 22290-180 Rio de Janeiro, RJ, Brasil}\\
${}^c${\em Departamento de F\'{\i}sica Matem\'{a}tica}\\
{\em Faculdade de Filosofia, Ci\^{e}ncias e Letras de Ribeir\~{a}o Preto}\\
{\em Universidade de S\~ao Paulo, Av. Bandeirantes 3900 
Ribeir\~{a}o Preto 14040-901, SP, Brasil}}
\end{center}

\begin{abstract}
In this letter we propose a new ansatz for the thermal string in the TFD formulation. From it, we derive the thermal vacuum for the closed bosonic string and calculate the thermal partition function in the blackhole $AdS$ background in the first order of the perturbative quantization. 
\end{abstract}

\newpage

\section{Introduction}

In previous papers, we started to formulate the microscopic properties of bosonic strings and $D$-branes in the perturbative limit of string theory in the Thermo Field Dynamics (TFD) formalism \cite{ivv1,ivv2,ivv3,ivv4,ivv5,ivv6,ivv7}. The main advantage of employing TFD is the possibility of explicitely constructing string and boundary states at finite temperature. In particular, one can obtain a description of thermal $D$-branes as states in the Hilbert space of closed thermal strings similar to the $D$-branes at zero temperature. However, most of the 
results obtained in these works only apply to bosonic strings in flat space-time and in the presence of the constant Kalb-Ramond field. The first extention of the formalism to GS superstrings and a different
($pp$-wave) background was done in \cite{alg1}(see also \cite{alg2,alg3,alg4,alg5,alg6} for general $SU(1,1)$ formulation of TFD applied to string theory). In \cite{alg3} the bridge between
the TFD and the Imaginary Time Formalism for the case of closed bosonic string was made. 
However, the study of string physics in more general backgrounds represents an imporatant task for phenomenological applications as well as for the development of string theory. 

String theory in the balck-hole $AdS$ background \cite{btz} was perturbatively quantized for the first time in \cite{ns1} and various properties of this system were analyzed in \cite{ns2,ns3,ns4,ns5,ns6}. The perturbative method employed in the study of the string dynamics in general curved spacetime was developed in \cite{ns7,ns8}. This method is particularly suitable to the construction of the thermal string in the TFD approach since in the first order approximation it provides the required canonical structure. The aim of this paper is to construct the perturbative string thermal vacuum in the presence of a black-hole $AdS$ spacetime. Also, we are going to clarify some aspects of the thermal string vacuum in TFD left open by the previous investigations. The thermal vacuum is crucial for the implementation of TFD and the construction of thermal states, since it contains the information about the Bogoliubov operators that map the theory to finite temperature and determination of the entropy operator.
Therefore, our results are important for the development of a systematic study of the microscopic properties of the boundary states of string theory in the blackhole $AdS$ background. 

The paper is organized as follows. In the following section we briefly review the perturbative canonical quantization of string theory in the blackhole $AdS$ background. In section 3 we construct the thermal vacuum
and compute the partition function at finite temperature. The last section is devoted to discussions.

\section{Zero Temperature Model}

The bosonic string in the blackhole $AdS$ background was quantized for the first time in \cite{ns1}. The canonical quantization was performed perturbatively with respect the natural dimensionless expansion parameter $\e$ which is related to the string tension and the inverse of the Hubble constant $\e = \a ' H^2$. The dynamics of the bosonic string in curved background is described by the following equations of motion \cite{ns7,ns8}
\be
\ddot{x}^{a} - x^{''a} + \G^{a}_{bc}
\left( \dot{x}^{b}\dot{x}^{c} - x^{'b}{x^{'c}} \right)=0,
\label{eqmotion}
\ee  
where $\G^{\mu}_{\r\s}$ are the Christofell symbols and $a, b, c, \ldots = 1,2,\ldots ,D+1$. The conformal symmetry of the worldsheet theory leads to the following set of constraints
\be
g_{ab}\dot{x}^{a}x^{'b} = g_{ab}\left(\dot{x}^{a}\dot{x}^{b} + x^{'a}x^{'b}\right)=0.
\label{constr}
\ee
In \cite{ns1,ns2} has been shown that the bosonic string in an arbitrary curved spacetime can be quantized perturbatively by expanding arround the exact solution of the equation of motion of its center of mass (the geodesic)
\be
x^{a}(\t ,\s ) = \sum_{n=0}^{\infty} \e^n \eta^{a}_{n}(\t , \s ),
\label{generalexpansion}
\ee
where $\eta^{a}_0 (\t )$ is a solution of geodezic equations and of the constraints
\bea
&~&\ddot{\eta}^{a}_{0} + \G^{a}_{bc}(\eta_0)\dot{\eta}^{b}_0\dot{\eta}^{c}_0 = 0,\label{geo1}\\
&~&g_{ab}(\eta_0)\dot{\eta}^{a}_0\dot{\eta}^{b}_0 = -m^2\a^{'2}.\label{geo2}
\eea
In order to quantize the string, one has to decompose the string coordinates in terms of transverse oscillations to the geodesics characterized by the following normal vectors $n^{a}_{\m}$
\bea
&~& g_{ab}(\eta_0 ) n^{a}_{\m} \dot{n}^{b}_{0} = 0, \label{normeq1}\\
&~& g_{ab}(\eta_0 ) n^{a}_{\m} {n}^{b}_{0} = \delta_{ab}. \label{normeq2}
\eea
In first order, the co-moving perturbations are given by the following relation
\be
\eta^{a}_{1} = \delta x^{\m} n^{a}_{\m},
\label{eqpert}
\ee
where $\m = 1, 2, \ldots , D-1$.
The equations of motion and the constraints can be easily solved for $\delta x^{\m}$ in the blackhole $AdS$ background and the solution has the following Fourier expansion 
\be
\delta x^{\m}(\t , \sigma ) = 
\sum_{n \neq 0} \sqrt{\frac{2n\Omega_n}{\a '}} \left[ \a^{\m}_{n}e^{-in(\Omega_n \t - \sigma)} + \b^{\m}_{n}e^{-in(\Omega_n \t + \sigma)} \right]
+ \sqrt{\frac{l}{2m}}\left[ \a^{\m}_{0} e^{-i\frac{m\a '}{l}\t} + \b^{\m}_{0} e^{i\frac{m\a '}{l}\t}\right].
\label{fouriersol}
\ee
Upon quantization, the Fourier coefficients become operators on the Hilbert space that satisfy the oscillator canonical commutation relations. The oscillator frequencies are $\omega_n = n \Omega_n$, $n = 1, 2, \ldots $, where we have used the following notation  
\be
\O_n = \sqrt{1+ \frac{m^2 {\a '}^2}{n^2 l^2}},
\label{notations}
\ee
where $l=1/H$.
The constraints in the blackhole $AdS$ spacetime take the form
\bea
\left( L^{\a}_0 - 2 \pi \a ' \right) \left| \Psi_{phys} \right\rangle = 0,
\label{leftconst}\\
\left( L^{\a}_0 - 2 \pi \a ' \right) \left| \Psi_{phys} \right\rangle = 0,
\label{rightconst}
\eea
where zero mode generators of the Virasoro algebra are given by the following relations
\bea
L^{\a}_0 &=& \pi \a ' \sum_{n>0} \frac{\o^2_n + n^2}{2\o_n}\sum_{\m=1}^{D-1}
\left( \a^{\m \dagger}_{n} \a^{\m}_{n} + \b^{\m \dagger}_{n} \b^{\m}_{n} - 2 \right)
+ \pi \a ' \sum_{n>0} \left( \b^{\m \dagger}_{n} \b^{\m}_{n} - \a^{\m \dagger}_{n} \a^{\m}_{n} \right) - \frac{\pi}{2}m^2 {\a '}^2,
\nonumber\\
L^{\b}_0 &=& \pi \a ' \sum_{n>0} \frac{\o^2_{n} + n^2}{2\o_n}\sum_{\m=1}^{D-1}
\left( \a^{\m \dagger}_{n} \a^{\m}_{n} + \b^{\m \dagger}_{n} \b^{\m}_{n} - 2 \right)
+ \pi \a ' \sum_{n>0} \left( \a^{\m \dagger}_{n} \a^{\m}_{n} - \b^{\m \dagger}_{n} \b^{\m}_{n} \right) - \frac{\pi}{2}m^2 {\a '}^2.
\nonumber
\eea
Here, the superscript indices $\a$ and $\b$ indicate the left- and right-moving quantities, respectively. As in the flat spacetime, one can construct the Hamiltonian of string oscillations $\hat{H}_{osc} = L^{\a}_0 + L^{\b}_0$ and the momentum $\hat{P} = L^{\a}_0 - L^{\b}_0$. To
obtain the full string Hamiltonian, one has to add the Hamiltonian of center of mass in the presence of the gravitational field
$H_{cm}$ to $H_{osc}$. The level matching condition should be imposed on the Hilbert space of the closed string in order to project the dynamics in the physical subspace  
\be
\pi \a ' \sum_{n>0} n \sum_{\m = 1}^{D-1} \left(\b^{\m \dagger}_{n} \b^{\m}_{n} - \a^{\m \dagger}_{n} \a^{\m}_n \right)\left| \Psi_{phys} \right \rangle = 0.
\label{levelmatching}
\ee
As in the flat spacetime, the physical states can be organized as eigenstates of the mass operator which is obtained from the constraints (\ref{leftconst}) and (\ref{rightconst}). For more details on this construction we refer the reader to \cite{ns1}.

\section{Thermal String Vacuum}

Let us construct the thermal vacuum of the bosonic closed string. According to the previous section, the total Hilbert 
space $\HH_{tot}$ in the first order perturbative quantization is the tensor product of the Hilbert spaces of transverse string oscillators. However, the physical Hilbert space $\HH_{phys}$ is the subspace of $\HH_{tot}$ defined by the 
constraints (\ref{leftconst}), (\ref{rightconst}) and (\ref{levelmatching}). In particular, this suggests that the trace in the TFD ansatz be modified in order to incorporate the level matching condition. The simplest way to implement the constraint (\ref{levelmatching}) in to the ansatz that defines the thermal vacuum is to introduce it through a delta function. Therefore, we modify the TFD ansatz \cite{tu,um} to the following one
\be
\langle \hat{O} \rangle = Z^{-1}(\b_T )\mbox{Tr}\left[ \delta(\hat{P}=0) e^{-\b_T \hat{H}}\hat{O}\right]\equiv
\langle\langle 0(\b_T)|\hat{O}| 0(\b_T )\rangle\rangle ,
\label{modifiedansatz}
\ee
where $\b_T = (k_B T)^{-1}$, $k_B$ is the Boltzmann's constant and $T$ is the temperature.
The last equality in (\ref{modifiedansatz}) defines the thermal vacuum by an expectation value relation. Note that the trace from (\ref{modifiedansatz}) is taken over the full Hilbert space $H_{tot}$ but due to the constraint, the calculated value take into account only the contribution of the physical states. Following \cite{tu}  
one can expand the thermal vacuum in the basis of the Hilbert space of closed string 
\be
\left.\left| 0(\b_T) \right\rangle \right\rangle_{osc} =
\sum_{w}\sum_{z} f_{w,z}(\b_T ) \left| w \right\rangle \left| z \right\rangle,
\label{vacexp}
\ee
where $w$ and $z$ are multi-indices that stand for {\em all} left- and right-moving oscillator indices, respectively. The explicit form of the undetermined coefficients $f_{w,z}(\b_T)$ is obtained by computing explicitely the trace and the vacuum expectation value from equation (\ref{modifiedansatz}). Let us introduce the following notations for the left- and right-moving number operators of the $n$th oscillator
\be
N^{\a}_n = n\sum_{\m=1}^{D-1}k^{\m}_n~~\mbox{and}~~N^{\b}_n = n\sum_{\m=1}^{D-1}s^{\m}_n,
\label{numberop}
\ee
where $k^{\m}_n$ and $\s^{\m}_n$ are natural number that denote the number of string excitations of the $n$th oscillator in the $\m$th spacetime direction in the corresponding string sector. By using the following analytic formula for the delta function
\be
\delta(\hat{P}=0)\equiv\delta(\hat{N}^{\a} - \hat{N}^{\b}) = \int_{-1/2}^{+1/2} ds
e^{2\pi i s\left(\hat{N}^{\a} - \hat{N}^{\b}\right)},
\label{deltaanalytic}
\ee
one can prove that the coefficients $f_{w,z}(\b_T )$ must satisfy the following consistency relation
\bea
&~&f^*_{w',z'}(\b_T)f_{w,z}(\b_T) = \nonumber\\
&~&Z^{-1}(\b_T)\bar{\delta}(w',w)\bar{\delta}(z',z)e^{\b_T [2(D-1) - \pi m^2 {\a '}^2]}
\int_{-1/2}^{+1/2} ds e^{ \pi \a ' \sum_{n>0}\left( \l^{\b}_{n}N^{\b}_{n} - 
\l^{\a}_{n}N^{\a}_{n} \right)},
\label{coeffrel}
\eea
where $\bar{\delta}$ is a short hand notation for the product of delta functions for each pair of indices in the multi-indez and $\l^{\a}_{n}$ and $\l^{\b}_{n}$ are functions of $s$ and $\b_T$ given by the following relations
\be
\l^{\a}_{n} = 2\pi i s + \b_T \f{\o^2 + n^2}{\o_n}~~,~~ 
\l^{\b}_{n} = 2\pi i s - \b_T \f{\o^2 + n^2}{\o_n}.
\label{lambdas}
\ee
In computing the r. h. s. of (\ref{coeffrel}) we have used a ket thermal vacuum which is just the hermitian conjugate of the bra vector (\ref{vacexp}). With this choice, the coefficients $f_{w,z}(\b_T)$ should be identified with vectors from a copy of the Hilbert space \cite{tu}. However, the presence of the constraint implies that these vectors be tensored with Columbeau functionals \cite{c}. Had the thermal vacuum been expanded in the basis of $\HH_{phys}$ instead of $\HH_{tot}$, this problem would have been avoided. Indeed, in this case the constraint did not appear in the trace explicitely. This suggest that in order to avoid complications that arise from the presence of the delta function such as delta function square root, one should balance it and compute the vacuum expectation value on $\HH_{phys}$ instead of $\HH_{tot}$. However, identifying the physical subspace without a delta function for constraints can be a nasty task. To circumvent arround this problem one can try to introduce the constraint in all vacuum expectation value computations. One way to do that is to redefine the ket thermal vacuum such that the delta function be included. Thus, we are lead to the following proposal for the ket thermal vacuum vector
\bea
&~&\left.\left| 0(\b_T)\right\rangle\right\rangle = 
Z^{-1/2}(\b_T)e^{\b_T[\pi\a '(D-1)-\f{\pi {\a'}^2m^2}{2}]}\times
\nonumber\\
&~&\sum_{k^1_1 = 1}^{\infty}\hspace{-0.2cm}\cdots \hspace{-0.2cm}\sum_{s^{D-1}_{\infty} = 1}^{\infty}\hspace{-0.3cm}
e^{-\f{\b_T\pi\a '}{2}\sum_{n=1}^{\infty}\sum_{\m=1}^{D-1}\g_n n (k^{\m}_n + s^{\m}_n)}
\left|k^1_1\cdots s^{D-1}_{\infty}\right\rangle
\widetilde{\left|k^1_1\cdots s^{D-1}_{\infty}\right\rangle},
\label{thermalvacuumket}
\eea
and for the bra thermal vacumm
\bea 
&~&\left\langle\left\langle 0(\b_T)\right|\right. = 
Z^{-1/2}(\b_T)e^{\b_T[\pi\a '(D-1)-\f{\pi {\a'}^2m^2}{2}]}\times
\nonumber\\
&~&\int_{-1/2}^{+1/2}\hspace{-0.3cm} ds \sum_{k^1_1 = 1}^{\infty}\hspace{-0.3cm}\cdots \hspace{-0.35cm}\sum_{s^{D-1}_{\infty} = 1}^{\infty}\hspace{-0.3cm}
e^{\sum_{n=1}^{\infty}\sum_{\m=1}^{D-1}\o^{\b_T}_n k^{\m}_n}
e^{-\sum_{n=1}^{\infty}\sum_{\m=1}^{D-1}\o^{\b_T}_n  s^{\m}_n}
\left\langle k^1_1\cdots s^{D-1}_{\infty}\right|
\widetilde{\left\langle k^1_1\cdots s^{D-1}_{\infty}\right|}.
\label{thermalvacuumbra}
\eea
Here, we have introduced the following short hand notations
\bea
\g_n &=& \frac{\o^2_n + n^2}{\o_n}\label{gamma}\\
\o^{\a}_n(s;\b_T)=(2\pi i s + \f{\b_T \pi \a '}{2}\g_n)n~~&,&~~
\o^{\b}_n(s;\b_T)=(2\pi i s - \f{\b_T \pi \a '}{2}\g_n)n
\label{os}
\eea
In (\ref{thermalvacuumbra}) and (\ref{thermalvacuumket}) we have explicitely written the 
multi-indices and the multi-indices sums. The tilde denotes the vectors from the identical copy of $\HH_{phys}$. By computing the vacuum expectation value of any observable of string in the thermal vacuum given by (\ref{thermalvacuumbra}) and (\ref{thermalvacuumket}) we automatically do the calculation on the physical subspace only, which is the tensor product
\be
\HH(\b_T)=\HH_{phys} \otimes \widetilde{\HH_{phys}}.
\label{thermalphsp}
\ee
The partition function can be calculated by imposing the normalization condition on the thermal vacuum. By using (\ref{thermalvacuumket}) and (\ref{thermalvacuumbra}) one obtains the following result
\be
Z(\b_T)=\int_{-\f{1}{2}}^{\f{1}{2}}\hspace{-0.3cm}ds\prod_{n=1}^{\infty}
\left[\frac{e^{2\b_T \pi \a '[(D-1)-\f{\a'm^2}{2} -\left(\f{\o^2_n + n^2}{\o_n}\right)]}}{2+ 
2e^{-2\b_T\pi\a ' \left( \f{\o^2_n + n^2}{\o_n} \right)} -
e^{-2\b_T\pi\a ' \left( \f{\o^2_n + n^2}{\o_n} \right)}\sinh(2\pi i s n)}\right]^{D-1}.
\label{partitionfunction}
\ee
The exact form of the integral in $s$ can be calculated but the result is not illuminating.
The relations (\ref{thermalvacuumket}), (\ref{thermalvacuumbra}) and (\ref{partitionfunction}) determine completely 
the thermal vacuum state as a state from the tensor product of the physical Hilbert space its (identical) tilde copy.
Note that if the computation is performed in the physical subspace the formula (\ref{partitionfunction}) must be replaced by one which does not depend on the integration. Nevertheless, the partition function as a function of temperature is the same.  

\section{Discussion}

The relations (\ref{thermalvacuumket}) and (\ref{thermalvacuumbra}) show that the effective frequencies of the string oscillation modes are given by the following relation
\be
\sigma_n = \pi\a '\left(\f{\o^2_n +n^2}{\o_n}\right).
\label{frequenies}
\ee
Thus, in the physical subspace, the partition function is given by the product of an infinite number of left- and right-moving oscillators which obey the Einstein-Bose statistics. This shows that one can construct the Bogoliubov operators for strings in blackhole $AdS$ background in the same way as in the flat spacetime in the first order in perturbation quantization. With the help of Bogoliubov operators we can investigate the algebraic structure of the theory at finite temperature, in particular the constructruction of the thermal string states, the boundary states and the calculation of their entropy \cite{ei}.

To conclude, in this paper we have proposed a modification of the ansatz for the calculation of the trace in TFD formulation of string theory which takes into account the constraints explicitely. A similar proposal was made in \cite{alg3} for the calculation of the one loop partition function. By using this new ansatz, we have constructed the thermal vacuum which has a delta function in the bra vector. This allows to compute the vacum expectation value from contributions of the vectors from the same physical subspace of the Hilbert space as the vectors that enter into the calculation of the trace. In particular, we have calculated the thermal partition function. The next step is to construct the thermal string states and boundary states and to derive their microscopic properties. We hope to report on this subject elsewhere \cite{ei}  

{\bf Acknowledgments}
I. V. V. would like to thank to S. Alves and J. A. Helay\"'{e}l-Neto for hospitality at CBPF where part of this paper was done. This work was supported by the FAPESP Grant 02/05327-3.

\end{document}